# Exploring self-driving labs for optoelectronic materials

Jonathan Staaf Scragg[1,2]

[1] Division of Solar Cell Technology, Department of Materials Science and Engineering, Uppsala University, 752 37 Uppsala, Sweden
[2] Wallenberg Initiative Materials Science for Sustainability, Department of Materials Science and Engineering, Uppsala University, 752 37 Uppsala, Sweden

**Abstract**

Self-driving laboratories (SDLs), by combining automation with machine learning-guided experiment selection, have the potential to transform experimental mateirals science. To date, most SDLs have been optimisation-driven, designed to rapidly converge on performance metrics, by embedding multiple mechanistic layers within platform-specific surrogate models. Such approaches excel at process tuning yet offer limited insight into the underlying physics governing synthesis-property relationships. Here we articulate a complementary paradigm: the exploration-driven, or scientific, SDL, whose primary purpose is the generation of data for data-driven science. We exemplify this concept for the case of inorganic optoelectronic materials, arguing that defect physics, which forms the central mechanistic link between synthesis conditions and functional properties, provides the foundation for designing a suitable SDL. Because defect populations and their spatial organisation cannot generally be resolved directly – nor fully predicted from first principles – the task of the SDL is to generate datasets in which thermodynamic and kinetic synthesis variables are systematically perturbed and defect-sensitive observables measured in parallel. From this basis, we propose a set of design principles for scientific SDLs that will enable them to operate "close to the physics" of optoelectronic materials, thereby generating transferrable and reusable datasets offering radical insight. We use $Cu_2ZnSn(S,Se)_4$ as a case study, both to show the scale of the task of defect-aware materials exploration as well to highlight as the deficiencies in the current paradigm. We propose that properly designed SDLs can generate the structured datasets necessary to enable mechanistic inference and advance synthesis-aware materials design.

## Introduction

Self-driving laboratories (SDLs) are emerging as a powerful new paradigm for materials research. By integrating automated sample synthesis and characterisation with machine learning (ML), SDLs enable high-throughput experimental campaigns to be carried out with minimal human intervention.[1] Typical SDL implementations employ active (sequential) ML approaches, most commonly Bayesian optimisation, to guide the selection of experimental conditions.[2] The SDL uses its measurements to iteratively train a surrogate model of the experiment (for example, a Gaussian process), which aids identification of conditions expected to yield the greatest information gain toward some goal. These are then implemented in the next experiment. In this way, the experimental space can be covered far more efficiently than would be possible using conventional approaches, especially when navigating large parameter spaces. Because their throughput can dramatically exceed that of manual experimentation, SDLs can rapidly accumulate large datasets, and thus form a natural component in the broader effort of data-driven science.

**Optimisation-driven vs exploration-driven SDLs**

The extent to which SDL-generated surrogate models and datasets can support *scientific discovery* – in the sense of yielding mechanistic insight, transferable understanding, and predictive models that extrapolate beyond their training data – depends on how SDLs are designed and operated. We distinguish two categories of SDL, illustrated in **Figure 1**, which embody fundamentally different philosophies. The first, and the most common, can be described as an "optimisation-driven" SDL. Such SDLs function as powerful automated process-engineering tools, prioritising rapid convergence figures of merit (FOMs), such as device efficiency[3], catalytic activity[4], reaction yield[5] or charge capacity,[6] by controlling the experimental inputs – typically process-specific parameters e.g., reactor flow rates, heater settings or source powers. Optimisation-driven SDLs can be extremely effective in experimental situations where maximising performance within a complex process-parameter space is the primary objective. Yet, their surrogate models necessarily collapse multiple mechanistic layers into a single mapping: the relationships between setpoints and actual experimental conditions, between synthesis conditions and materials formation processes, between structural and chemical material properties and functional performance, and finally between these and the measured FOMs. While this collapse is central to their effectiveness as optimisers, it renders the resulting model platform-specific, with limited ability to extrapolate beyond its training data.[7]

  We therefore advocate focus on a second, less-developed category of SDL, which we refer to as the "exploration-driven" or "scientific" SDL. Here, the objective is not optimisation, but the generation of rich datasets capturing the materials physics and chemistry of the system under study as explicitly as possible. The focus is not on solving a proximate process engineering problem, but rather on accumulating data as the foundation for data-driven scientific progress over time. In part, this distinction relates to the exploitation/exploration trade-off familiar from active learning methods: the scientific SDL's control algorithms must deliberately explore *non-optimal* regions of experimental space, including approaching phase boundaries and traversing distinct kinetic regimes, to capture the full structure of the composition-synthesis-property relationships. But the distinction goes beyond this too, because it influences the very design of the experimental setup. To serve their function, scientific SDLs must operate "close to the physics" of the system under investigation, which implies several things: (1) basing SDL design and its parameter space closely on the physics and chemistry of the materials at hand. This is in order to probe *all* relevant axes and to, as far as possible, enable different physical effects to be separated in the experiment (2) that actual synthesis conditions – rather than instrument-specific setpoints – must be known, through appropriate sensing, calibration or process modelling, (3) that SDL measurements should probe fundamental material properties, rather than figures of merit. Scientific SDLs will require greater sample throughput than their optimiser cousins, since exploration inherently requires more extensive sampling of the experimental space.[2] Therefore, scientific SDLs will need more sophisticated search algorithms – physics-informed ML or hybrids of ML and mechanistic modelling – to constrain learning to physically plausible regimes. Additionally, advanced experimental design will be needed to increase sample throughput.

If these goals can be met, the surrogate models and datasets produced by scientific SDLs will encode synthesis-property relationships that are more complete and more closely aligned with the underlying physical processes in the materials at hand. Such datasets can then form the foundation for (a) deriving mechanistic understanding of synthesis–property relationships rather than merely identifying correlations, (b) developing transferable, physically grounded models of materials growth and function, and (c) training higher-level ML models (for example, physics-informed neural networks) capable of predicting outcomes beyond the materials and processes tested. This is the most realistic path toward *inverse design* – the capability to predict materials and processes that will deliver required properties.[8]

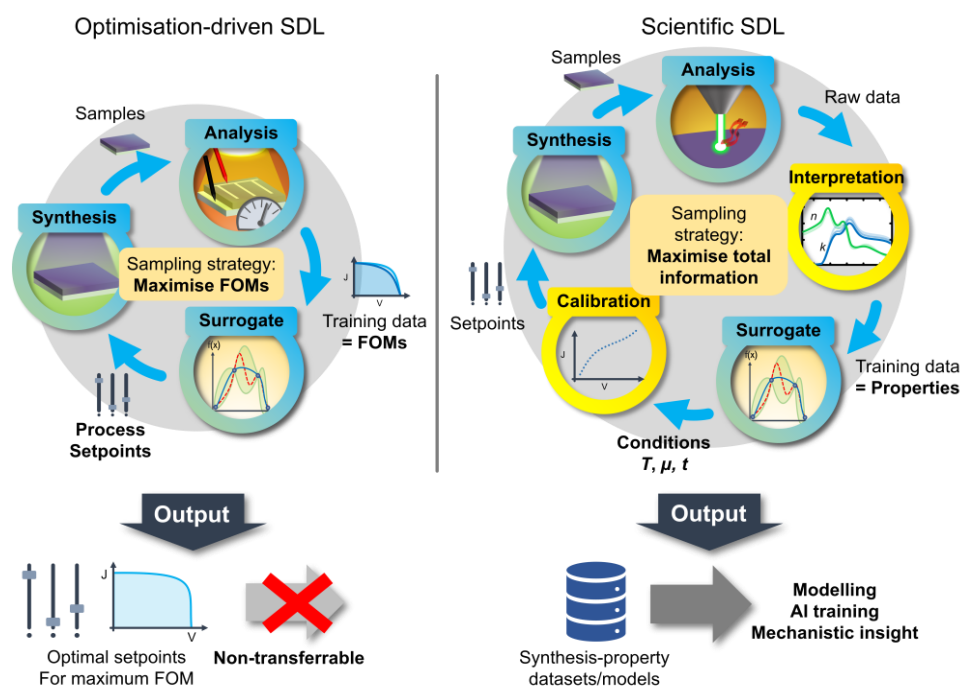

**Figure 1:** Contrasting roles of self-driving labs (SDLs) in materials science. **Left:** "optimisation-driven" SDL: application-oriented system prioritising performance maximisation in a device context. A machine learning surrogate is trained to predict figures of merit (FOMs) for given synthesis setpoints. **Right:** "scientific" SDL, prioritising generation of reusable datasets capturing materials physics/chemistry. The surrogate is trained to predict materials properties based on synthesis conditions.

**SDLs for inorganic optoelectronics**

Our focus here is on scientific SDLs for inorganic optoelectronic materials. Optoelectronic materials are a central pillar of modern technology, underpinning information and communication technologies, energy conversion and efficiency, sensing and imaging, and emerging applications such as photocatalysis and quantum photonics. While silicon, III–V and II–VI semiconductors, wide-bandgap nitrides and oxides, and more recently halide perovskites already support huge markets, they occupy only a small fraction of the accessible materials design space.[9,10] Discovering new optoelectronic materials with tailored band gaps, carrier transport, stability, defect tolerance and earth abundance remains a strategically important task.

Scientifically, the biggest challenge for understanding and controlling optoelectronic materials lies in their defect physics and chemistry. Variations in defect populations and spatial arrangements can modify band gaps and band edges, influence optical selection rules and luminescence, set free-carrier concentrations through doping, scatter carriers and phonons, and alter diffusion pathways for mass transport.[11] Although defects are often associated with degraded performance – for example through band tailing or non-radiative recombination – controlled defect populations are essential in many technologically important materials [12–15]. As complexity increases from binary to ternary, quaternary, and pentanary systems, the space of possible defects and their spatial organisation grows combinatorially, rendering the internal state of the crystal highly sensitive to the synthesis pathway.

At the same time, the complexity of defect populations and their constellations – spanning length scales from point defects to extended aggregates and disorder – makes them difficult to analyse using either first-principles theory or experimental characterisation alone [14,16,17]. This difficulty has long hindered progress toward a predictive understanding of synthesis–property relationships in crystalline functional materials, for optoelectronics or otherwise.

We argue that scientific SDLs and data-driven approaches offer the most promising path forward, by extending existing strategies for indirect defect inference. One such strategy is "high-throughput" materials characterisation, which often makes use of combinatorial thin-film samples, where functional properties can be mapped as a function of composition on single substrates.[18] In such a context, observed property trends can help identify candidates for the underlying defects: whichever defect caused the change in properties must also be consistent with the change in composition (or chemical potential, assuming the material remains single-phase). In essence, observing such trends provide *constraints* on the possible defect types, from among a large number of possibilities. However, composition is only one of many dimensions influencing defect populations and configurations, and by itself does not fully constrain the possibilities. For example, the same change might be explained by a substitutional defect, or a vacancy, or a complex of both.

Let us now envisage an SDL that can produce *many* such combinatorial samples, in which besides composition we also vary other synthesis parameters spanning the thermodynamic and kinetic axes that control defect populations. Using this SDL, we could build datasets containing highly detailed property trends – creating a rich system of constraints on possible defect types and their arrangements. Such datasets would form a powerful asset for evaluating models of defects in crystalline materials, whether such models are based on physics, on heuristics or – perhaps most likely – proposed by generative AI. This, we argue, is the core task of a scientific SDL for optoelectronic materials: to generate reality-grounded datasets with sufficient breadth, structure, and physical basis to enable evaluation of defect models and, ultimately, to support rational materials design.

In this paper, we describe an approach to design and operate such an SDL. To elucidate the experimental parameters the SDL should control, we shortly refer to the theory of point defects in inorganic materials and how this plays into real synthesis. We introduce the concept of the *defectome* of a material, as a collective descriptor encompassing defect populations and their spatial organisation across length scales, which provides a convenient language for discussing defect evolution with synthesis. We articulate some design principles that bring the scientific SDL "close to the physics" of optoelectronic materials, and which lead to definition of the complete experimental space for the SDL operations. We exemplify some of the ideas using $Cu_2ZnSn(S,Se)_4$ (CZTSSe) as a case study, which also demonstrates the deficiency in current experimental paradigms for studying complex materials. We conclude by describing our own progress toward implementing such an exploratory SDL at Uppsala University.

## Theory
### Basic theory of dilute defects in inorganic crystals

The conventional theoretical description of defects in crystalline solids is built around isolated point defects in the dilute limit and their formation energetics.[11,17] In first-principles treatments, the formation energy of a defect species $D^q$ in charge state q is typically written at T=0 K as

$$\Delta H_f(D^q) = E_{tot}(D^q) - E_{tot}(bulk) - \sum_i n_i \mu_i + q(E_F + E_V) + E_{corr}$$

This expression establishes a simple equilibrium picture in which each atomic species *i* exchanges atoms with a thermodynamic reservoir characterised by a chemical potential $\mu_i$. The defect formation energy then determines, via Boltzmann statistics, an idealised equilibrium defect concentration,

$$N(D^q) \propto exp\left(-\frac{\Delta H_f(D_q)}{k_B T}\right)$$

Varying the chemical potentials $\mu_i$ moves the system through a multidimensional chemical-potential space, which in computational studies functions as an abstract representation of synthesis conditions. The allowable range of $\mu_i$ is bounded by competition with other phases that may form from the same components. This defines an envelope of $\mu_i$ values within which a given phase is stable. This envelope is illustrated using the example of $Cu_2ZnSnS_4$ (CZTS) in Figure 2(a).[19] Using this framework, one can identify which point defects are energetically favoured under different chemical environments, and how their concentrations vary across the stability region. Changes in $N(D^q)$ driven by shifts in $\mu_i$ constitute the most basic origin of synthesis-property relationships.

To go to finite temperatures, where synthesis necessarily occurs, we must replace the formation enthalpy $\Delta H_f$ with a temperature-dependent free energy $\Delta G_f(T)$. This introduces additional entropic and other vibration entropy contributions to the defect formation energy, both directly as well as via changes in the chemical potentials of the reservoirs.[11] At this point, we already touch the limits of what first principles calculations can feasibly compute, though we are still far from a realistic scenario.

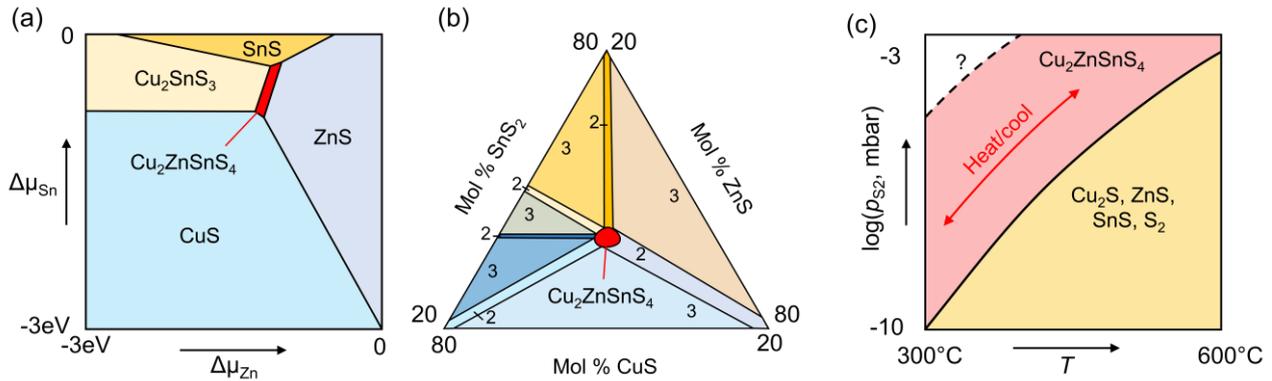

**Figure 2.** Visualisations of single-phase regimes using example of $Cu_2ZnSnS_4$ (CZTS): (a) Envelope of chemical potentials $\mu_i$ within which CZTS is the stable phase in equilibrium calculates (at 0K), adapted from.[19] (b) Phase diagram section in compositional space indicating the experimental single-phase region of CZTS at 400°C (central, red shaded region). Outside this region, numbers indicate the number of co-existing phases. Adapted from.[20] (c) Phase diagram in temperature-partial pressure space. The lower bound, below which S is lost and binary phases result, is known approximately from literature. A possible upper bound, where over-sulfurization may result in breakdown to simpler phases, is indicated, marked '?'. Adapted from.[21]

**Beyond the dilute limit – introducing the defectome**

The next increase in realism arises once defect–defect interactions are considered. These interactions are especially prevalent in multinary materials, such as those of interest for optoelectronic applications, where the number of distinct point defects can be large and often includes multiple low-energy, high-concentration species. Interactions mediated by electrostatics, elastic strain, and local chemical environment modify defect free energies through binding contributions and introduce configurational entropy associated with the many possible spatial arrangements. As a result, defect pairs and complexes, extended clusters, line defects (e.g., dislocations), planar defects (e.g., twins and grain boundaries), and even bulk disorder can all emerge as equilibrium or metastable outcomes at finite temperature.[16,22–26] The consequence is a vast proliferation of possible internal states of the crystal, far exceeding what can be described by isolated point defects and small supercells.

To refer collectively to this rich variability in defect constituents, we introduce the term "defectome" of a material. The defectome, $\mathcal{D}$, denotes the collected populations of defects and their spatial organisation across relevant length scales in a given state of the material. While we do not propose a definitive mathematical definition, a minimal conceptual representation would be a state vector such as:

$$\mathcal{D} = \left(c^{(1)}, c^{(2)}, \ldots; g^{(1)}, g^{(2)}, \ldots; l_{disl}, A_{GB}, A_{plan}, \ldots \right)$$

where the components represent concentrations of possible point defects (including charge states, polarons etc) $c^{(i)}$, defect complexes $g^{(j)}$, line-defect densities ($l_{disl}$), and areal densities of planar defects and grain boundaries ($A_{plan}$, $A_{GB}$). For any set of equilibrium thermodynamic conditions, and indeed any specific synthesis path, the components of the defectome will attain certain values, representing different possible internal states of the material. Thus, while the materials *genome*[9] refers to the external design space of all possible materials, the *defectome* refers to the internal design space of states that may be adopted by a specific material. Example defectome components are indicated in **Figure 3.** Introducing the defectome as a concept is useful because it emphasises that synthesis-property relationships are governed not by a single dominant defect, but by the collective behaviour of many interrelated and interacting defect populations and motifs, each of which may contribute differently to materials properties. In particular, these important effects span a continuum between the atomic – point defects – and the macroscale – microstructure – prompting the use of a new terminology that encompasses states between these extremes.

A properly defined defectome object might become a useful way to store information about materials, e.g., results from first principles calculations or dynamic simulations, in a compact form that is compatible with machine-learning workflows, or be used in development of mechanistic models of defect chemistry, analogous to reaction-diffusion frameworks in other fields.[27] A valuable community discussion could be had around the utility and proper definition of defectomes in optoelectronic materials, however here we use *defectome* only as a linguistic device, with any change in population or configuration of defects, from point defects to extended defect structures, being referred to as *changes in defectome state*.

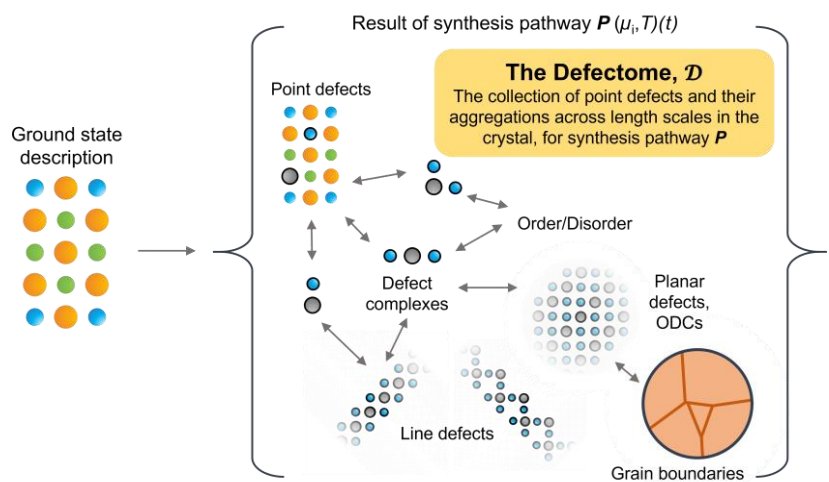

**Figure 3:** Representation of components of the defectome of a crystalline material - the defectome is defined by the collection of different point defects and their aggregations across different length scales. Each synthesis path *P* will in general lead to a different defectome state due to variations in the balance of formation energies and entropies as well as due to kinetics of point defect migration.

**Toward experimental synthesis**

When we move from equilibrium theory to real synthesis, the picture evolves further. In experiments, we do not control chemical potentials. Instead, we vary compositions $C_i$ and partial pressures $p_i$ (or arrival rates) of the components. Rather than an envelope in $\mu_i$-space, there is one in $\{C_i, p_i, T\}$-space within which a given phase is stable without the presence of secondary phases. We refer to this envelope as the single-phase region. The single-phase region determines the limits of a material's intrinsic synthesis-property relationships. For conditions within it, the defectome varies directly due to changes in the effective $\mu_i$ values. Outside, the formation of secondary phases pins chemical potentials – and thus defect populations – along tie-lines or tie-triangles. Correspondingly, property changes are dominated by phase coexistence, rather than further defect evolution. Experimental single-phase regions in both composition and partial-pressure/temperature spaces are shown for CZTS in Figure 2(b) and (c).

Finally, the defectome of a material cannot respond instantaneously to changing conditions that might occur during synthesis and post-growth processing. Time-dependent processes include spontaneous creation and annihilation of defects (relatively fast) and *migration* of point defects, which enables aggregation, reordering and dissociation at a various time and length scales, including processes such annealing and grain growth.[27] The degree to which the defectome can attain equilibrium during synthesis will depend closely on the temperature-time profile which is applied; fast cooling, for example, will freeze in high-temperature metastable defectome states, while slower cooling will allow the defectome to track the changing equilibrium state more closely. At the microscopic level, defect migration proceeds via thermally activated hops between lattice sites, characterised by migration barriers $\Delta E_m$. These barriers are themselves modified by the local defect environment, through changes in potential well depths and available migration pathways – this is why vacancies, interstitials, and extended defects such as grain boundaries dramatically enhance solid-state diffusion.[11] As a result, the kinetics of defectome evolution will strongly depend on its current state as well as on the processing trajectory.

The thermodynamic and kinetic considerations of defectome formation and evolution provide key guidance if we wish to probe defectomes in inorganic materials over a large possibility space and capture the resulting property trends. Specifically, our SDL must be able to make controlled variations in *composition* and *partial pressure* of all relevant components, as well as probe many different *temperature-time profiles* (temperature, duration, cooling rates), to capture processes occurring in different kinetic regimes. Co-variance of these parameters is critical, due to the fact that the defectome mediates its own evolution by setting defect-migration conditions ($\Delta E_m$).

## Results and Discussion

We now make proposals for design and operation of a scientific SDL for optoelectronic materials, using the defect theory as a guide. We are not aiming to specify all design choices, but rather a set of guiding principles. A diversity of different solutions to these principles is encouraged in order to discover the best possible outcomes. We note that while optoelectronics is in focus, the same considerations apply to other multinary inorganic functional materials. We first give a basic overview of the SDL workflow, then focus on how to structure experiments so that the defectome state is perturbed along physically motivated axes, and how to define and constrain the relevant experimental space. We use $Cu_2ZnSn(S,Se)_4$ (CZTSSe) to illustrate the scale of the defectome-exploration challenge and the potential of SDL-based methods, and report briefly on our own nascent SDL.

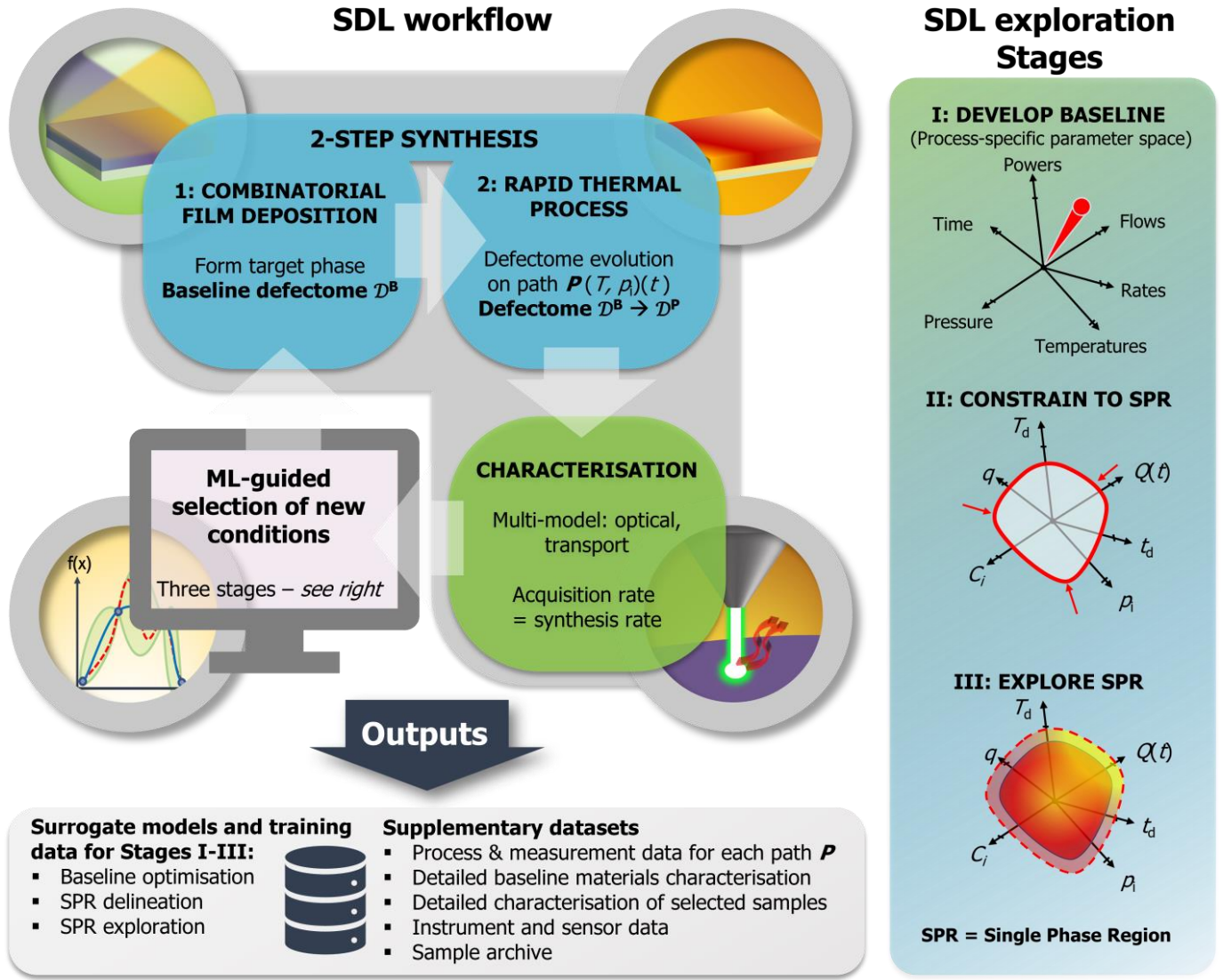

**Figure 4** Overview of the workflow of a "defect aware" SDL for optoelectronic materials, including two-step synthesis and characterisation and three Stages (I-III) of SDL exploration, according to the design principles described in the text. Also indicated are the main outputs of the SDL in terms of dataset components and samples.

**Design principles for exploratory SDLs in optoelectronics**

An overview of the proposed scientific SDL is given in **Figure 4,** which will be referred to throughout this section. In operation, it will synthesise combinatorial thin film samples based on experimental parameters chosen by a machine learning (ML) model, it will measure some properties of the resulting samples, and the results will be used to improve the model in an iterative fashion. Since the focus is primarily on workflow design, specific ML models will not be discussed here. The SDL in **Figure 4** uses a two-step synthesis process and has three phases of operation with different ML targets. The design and operation choices illustrated in

Figure 4 are motivated with reference to four design principles, that enable the SDL to efficiently generate datasets that are "close to the physics" of synthesis-property relationships in optoelectronic materials.

**Design principle 1: Function-first combinatorial samples as the "unit experiment"**

The SDL described here is an evolution of existing high-throughput approaches, making combinatorial deposition and characterisation the basic "unit experiment". Using thin films as the medium for materials exploration is extremely material-efficient compared to e.g., solid state synthesis. Combinatorial thin films have composition gradients across their area, induced during deposition by a specific source geometry; this allows the all-important composition space to be compressed onto single substrates. Suitable methods include physical vapor deposition (PVD) - sputtering, thermal evaporation, molecular beam epitaxy or pulsed laser deposition - but also solution coating, e.g., inkjet printing of nanoparticles.[28] Given that our SDL should generate transferrable, accurate datasets, it is important to focus on material quality: purity, uniform morphology etc, to allow for proper interpretation of characterisation data. The SDL will also have to make *many* samples, making optimisation for high-deposition rate a key priority.

In terms of characterisation, our SDL needs fast and automatable techniques in its main loop. For combinatorial samples, spatially resolved measurements are necessary – imaging (ideal), or point-by-point mapping using gridded or active-learning sampling strategies. The slowest stage in the SDL will be rate-limiting, which means that the turnaround time for the synthesis and characterisation stages should be approximately equal. Given the limited time frame for characterisation, the SDL must prioritise measuring *functional* properties – in this case, optical response and charge transport – over structural and chemical ones. This is because functional properties are much more sensitive to defects, and thus provide more tangible information to the SDL to navigate its experimental space. They can also be time-sensitive (due to degradation), so it is critical to acquire them directly after synthesis. Crystal structure, phase and composition metrics will change much less over the experimental space and over time, while also being slow to map on each sample. In principle, the SDL need *only* apply functional property characterisation to every sample, and route occasional samples to more detailed/slower characterisation – identified during or after completion of the SDL campaign.

**Table 1** lists a selection of possible defect-sensitive measurements. Imaging based methods such as hyperspectral optical or photoluminescence imaging are ideal for speed and resolution. Spot-based measurements or those that require time for signal accumulation will have to trade off speed against resolution and/or measurement noise. Ideally, several complementary measurements should be used to give access to multiple dimensions of information – this will help further constrain explanatory defect models.

| Method | Defect-Sensitive Features |
| --- | --- |
| Optical UV/vis T/R spectroscopy & Hyperspectral Imaging | Bandgap ($E_g$): Sensitive to composition/strain/disorder. Urbach Energy ($E_U$): Proxy for structural disorder & potential fluctuations. Sub-gap Absorption: Direct absorption by mid-gap defect states. Absorption coefficient and refractive indices: Bulk electronic structure. Microstructure (imaging); some microstructural features visible in optical images |
| Photoluminescence (PL) mapping or imaging | Intensity: Non-radiative recombination rate (trap density), quasi-Fermi level splitting. FWHM: Crystal disorder /inhomogeneity. Peak Shift: Strain or alloying disorder. Sub-gap Tail (Urbach): Band-tail states / defect density. |
| Time-Resolved PL (TRPL) | Minority Carrier Lifetime: Direct proxy for bulk defect density and surface recombination velocity. Decay Curve Shape: Trap saturation or multi-level recombination. |
| Raman Spectroscopy | Peak FWHM: Phonon lifetime reduction due to structural disorder/defects. Peak Shift: Residual stress/strain. Forbidden Modes/relative intensity changes: Appearance indicates symmetry breaking/structural transitions |
| Terahertz Spectroscopy (OPTP/TRTS) | Conductivity/Mobility: Carrier scattering rates due to defects/grain boundaries. Lifetime: Ultra-fast trapping dynamics (<1 ns). |
| Microwave Conductivity (TRMC) | Peak Photoconductivity: Product of mobility and carrier generation (trap influenced). Decay Kinetics: Trapping/recombination rates. |
| Four-Point Probe (4PP) | Resistivity: Integrated metric of carrier scattering (mobility reduction due to defects/GBs) and carrier compensation (traps reducing active carriers). |

**Table 1:** SDL-compatible measurements of functional properties meeting the following criteria: (1) having relatively short measuring times, (2) being non-destructive, (3) having the possibility for spatial resolution (mapping/imaging), and (4) not requiring sample-preparation such as contacting.

**Design principle 2: Separate phase formation from defectome evolution**

The next principle constrains the synthesis methodology in the SDL, motivating the use of separate film deposition and thermal processing stages as seen in **Figure 4**. In growth of functional thin film materials, two processes occur: (1) formation of the target phase on a substrate, by deposition and reaction of components from initially separate sources, and (2) what we can call "defectome evolution" – meaning progression of the defectome state in the formed phase toward an equilibrium, which is dictated by the given synthesis conditions. This includes processes such as grain growth, since grain boundaries and other planar/line defects are regarded as components of the defectome, and also includes post-deposition steps such as cooling.

The motivation for separation is simple: the thermochemical space that affects the defectome is much larger than the subset of conditions under which film deposition and phase formation can occur. Consider for example PVD methods: each deposition platform imposes strong, technique-specific constraints: vacuum level, arrival rates, plasma chemistry, substrate heating, etc., because they must simultaneously deliver reactants *and* overcome kinetic barriers to phase formation. If defectome variation were only probed through PVD-compatible conditions, then the dataset would (a) become entangled with platform-specific transport physics and (b) be biased toward a narrow region of experimental space i.e., the intrinsically low-pressure, high temperature regime of PVD growth. In contrast, defectome changes can occur under conditions that are *not* growth-compatible, such as at high partial pressures or modest temperatures, or vary over conditions that are not easily controlled in a PVD system, such as cooling rates. Operationally, the separation of phase formation and defectome evolution is achieved using a two-step materials synthesis process:

(1) **Phase formation:** Deposit combinatorial films of the target phase in one dedicated system optimised for that purpose, such as a PVD chamber, providing energy (substrate heating/post annealing) to promote reaction and phase formation.
(2) **Defectome evolution:** Modify the defectome state of the material in a separate dedicated system capable of accessing the full experimental space for defectome evolution, e.g. a rapid thermal processor (RTP) with controlled atmosphere.

This division explicitly separates the "engineering" component (optimisation of deposition method) from the "scientific" component (exploration of defect physics), in line with the concept of a scientific SDL. To prevent the combined two-step parameter space from becoming unmanageable, Step (1) can be fixed. The SDL should conduct an initial stage of operations (Stage I in **Figure 4**) using only the deposition step, in which it aims to find suitable conditions, not for peak material performance, but for a *reproducible baseline process*, for preparation of films with uniform thickness, minimal stress/strain, controlled composition gradients and verified single-phase formation at the stoichiometric location. A stable baseline process will result in samples that always have the same defectome state, $\mathcal{D}^B$.

Having established this baseline, subsequent SDL operations use both sample preparation steps in series. The workflow now involves repeatedly synthesising baseline samples and transferring them to the rapid thermal processor (RTP), in which a different thermochemical path $P$ is applied to each sample, transforming the baseline defectome state $\mathcal{D}^B$ into a new state $\mathcal{D}^P$. This is repeated along many different paths, and functional properties of each resulting sample are measured to accrue a large dataset.

Because all paths begin from the same baseline starting point, differences in measured properties can be interpreted primarily in terms of defectome evolution rather than uncontrolled variation in phase formation or platform-specific growth processes. This makes the dataset close to the physics of interest, ensuring better transferability. Since $\mathcal{D}^B$ serves as the reference for the entire SDL campaign, the baseline material should be characterised and reported exceptionally deeply.

**Design principle 3: Setting the experimental space for defectome manipulation**

We now focus on defining the possible defectome evolution paths, $P$, and the associated experimental space for the SDL to probe in its RTP step. As highlighted in the Theory section, the core factors controlling the defectome state are thermodynamic and kinetic ones: temperature $T$, time $t$, and the effective chemical potentials set by the solid-state compositions $C_i$ and gas-phase partial pressures $p_i$ of each component $i$. From these we can define the axes of the experimental space to be explored.

First, the $p_i$ axes. The number of these depends on the equilibrium vapour pressures of relevant components above the target phase. If a component has a non-negligible vapour pressure, its concentration in the film – and the prevalence of related defects – will be sensitive to its partial pressure in the gas phase. Most anionic components such as O, S, Se, P, Cl and some molecular species e.g., SnS(e) and $Sb_2O_3$ fall into this "volatile" category. Each volatile component gives one $p_i$ axis to control. Accurate control and measurement of these partial pressures is challenging – especially for condensing species – *but essential*, motivating the use of continuous-flow RTP reactors with in situ gas analysis.

Next, the $C_i$ axes: the ingoing solid-state composition of the sample will be determined by the baseline film deposition step. The baseline process only needs to provide combinatorial gradients for the non-volatile components; since the volatile components will have their chemical potentials controlled separately via $p_i$. For $n$ non-volatile components in the material, we have $n-1$ independent $C_i$ axes, of which up to two may be included in a single combinatorial sample. Thus, ternary and quaternary materials can be explored from a single combinatorial baseline, if at least one of their elements is volatile. Higher order compounds will require multiple baseline processes to span the additional $C_i$ axes.

Finally, the $T$ and $t$ axes are defined though the achievable thermal profiles in the RTP. We use a parameterised profile, as in **Figure 5**. This involves rapid heating to a target temperature $T$, followed by a dwell for a time period $t$. To simplify the experiment, we choose a high heating rate (e.g., 20-100°C/s) to avoid significant defectome transformation during the heating process. The sample then undergoes controlled cooling at a rate $Q$. To enable the kinetics of defectome cooling to be investigated, we include a final parameter $q$, which is defined as the fractional value of $T_d$ at which the sample is quenched (cooled at the fastest possible rate). This thermal profile allows the SDL to perturb the defectome over a range of states between near-equilibrium and strongly-non-equilibrium situations, and introduces four additional experimental axes: $T$, $t$, $Q$, and $q$. Strictly, the $p_i$ values should also be ramped during the heating and cooling stages, to maintain a steady solid-gas equilibrium and avoid unintended phase transitions (illustrated in **Figure 2**(c)).

Based on the above, **Table 2** shows the number of independent experimental axes for defectome exploration in different optoelectronic candidates.[14,29–32] We see that in common cases, we have at least 5-6 independent experimental axes to vary, after discounting those contained in the combinatorial sample. This large experimental space underscores the importance of going beyond "composition-only" high-throughput strategies.

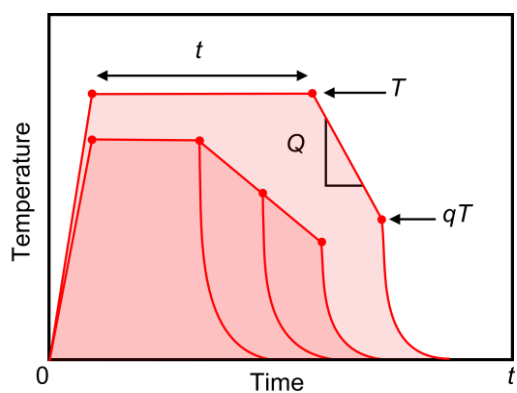

**Figure 5** Parameterised thermal profile for rapid thermal process in step two of sample synthesis (defectome evolution step), illustrating various profiles that can result.

| Compound | Non-volatile components | Volatile components | Independent $C_i$ axes | $C_i$ axes using combinatorial method | Independent $p_i$ axes | Thermo-kinetic axes[a] | Total axes |
|---|---|---|---|---|---|---|---|
| Cu(In,Ga)Se$_2$ | Cu, In, Ga | Se | 2 | 0 | 1 | 4 | 5 |
| Ba(Ti,Nb)O$_3$ | Ba, Ti, Nb | O | 2 | 0 | 1 | 4 | 5 |
| Cu$_3$PS$_4$ | Cu | P, S | 0 | 0 | 2 | 4 | 6 |
| (Ba,Sr)(Zr,Ti)S$_3$ | Ba, Sr, Zr, Ti | S | 3 | 1 | 1 | 4 | 6 |
| Cu$_2$ZnSn(S,Se)$_4$ | Cu, Zn[b], Sn | S, Se, SnS, SnSe | 2 | 0 | 4 | 4 | 8 |

**Table 2:** Experimental axes for defectome exploration in some example multinary compounds explored for optoelectronic applications. [a]$T$, $t$, $Q$, $q$. [b]Although elemental Zn is volatile, its partial pressure above CZTSSe is the relevant factor for defining a $p_i$ axis, and this is negligible.

**Design principle 4: the single-phase region constrains the experimental space**

To constrain this large experimental space, the natural next step is for the SDL to determine the boundary of the single-phase region of the target phase (see **Figure 2**(b), (c)), because this bounds the space for defectome manipulation. Although the obvious means to detect the single-phase region boundaries would be via measurements of phase content, it is also possible to detect them via functional property measurements of the target phase, which aligns with our "function-first" concept (Design Principle 1). This is possible because the multi-phase equilibria outside the single-phase region pin $\mu_i$ values (in other words, the relevant defects attain their maximum possible concentration), which means that defect-sensitive signals from our target phase will plateau as the phase boundary is crossed (or, alternatively vanish or demonstrate another discontinuity). This effect was demonstrated in detection of single-phase region boundaries for CZTS using Raman and PL mapping.[33,34] While less definitive than direct phase mapping, it will be worth investing effort in boundary-finding procedures following this concept, to increase the speed and reduce the complexity of the SDL.

Thus, with reference to **Figure 4**, Stage II in the SDL operations focusses on delineating the single-phase boundaries using functional property measurements. This can be driven using some boundary-finding ML algorithm that looks for the relevant discontinuities in properties across the experimental space, to define a closed region within it. Once this is done, the final Stage (III in **Figure 4**) can proceed, in which the SDL explores the variations in functional properties only within the single-phase region, applying an algorithm that is suited for mapping the expected smooth changes in its interior.

**The scale of the task: example of Cu$_2$ZnSn(S,Se)$_4$**

To illustrate the scale of the task of defectome mapping, we use the example of Cu$_2$ZnSn(S,Se)$_4$ (CZTSSe). CZTSSe and its end members CZTS and CZTSe are among the most intensively studied multinary semiconductors for thin-film photovoltaics, motivated by the desire to avoid scarce or toxic elements. Despite steady progress – including a recent record efficiency of 16.6%[35] – device performance remains well below that of competing technologies. A wide range of defect-mediated synthesis–property relationships has been reported, associated with disorder, off-stoichiometry, band-gap fluctuations, and deep defects. Yet, these effects remain incompletely quantified, their microscopic origins contested, and the best synthesis routes are empirically derived rather than designed.[36] **Table 3** gives a non-exhaustive selection of defectome components in CZTSSe. While some components exhibit distinctive behaviour, many have overlapping dependencies on synthesis parameters and affect similar properties. This overlap explains the difficulty in assigning causal relationships using conventional experimental approaches, and underlines the value of more systematic defectome mapping.

| Component | Means of identification | Influences on properties | Known/Expected dependence on synthesis |
|---|---|---|---|
| $Cu_{Zn}$, $V_{Cu}$ | DFT calculations[25] | Shallow acceptors → p-type doping | Composition-dependent formation energy – favoured in Cu-poor or Cu-rich/Zn-poor conditions |
| $Zn_{Cu}$ | DFT calculations[30] | Compensating shallow donor | Composition-dependent formation energy |
| Cu-Zn disorder/ antisite pairs | Direct detection by NMR,[37] indirect via Raman scattering.[38] | Band gap shift, potential fluctuations, band tailing | - Strong function of cooling rate below transition temperature (200-265°C). Ordering kinetics depend on composition due to vacancy/interstitial enhancement |
| Various defect complexes | Neutron diffraction,[25] DFT calculations[30] | Enable off-stoichiometry, band gap/potential fluctuations | Composition-dependence of concentration. Expected cooling-rate dependence on configuration/clustering |
| Sn(II), $V_S$ | DFT[39]; inferred from decomposition behaviour under S-poor conditions[21] | Deep defects – recombination centres – will affect e.g., PL yield and device voltage | Expected function of combined S and SnS pressures and cooling rate |
| Line and planar defects | TEM and DFT[23,24] | Carrier trapping – e.g., affecting mobility, possible recombination sites | Expected function of synthesis time, cooling rate, surface stability (i.e., S/SnS partial pressures) |

**Table 3:** Known or proposed components of the defectome of CZTSSe, including information on how they have been identified, which properties they influence, and how their populations are known or expected to depend on synthesis conditions.

Following the framework developed, we can define the experimental space required for defectome exploration in CZTSSe. As shown **in Table 2,** this space need contain no explicit $C_i$ axes, since the proportions of Cu, Zn and Sn can be independently varied on a single combinatorial sample (the single-phase region is about 1-5 at% wide between 400-900°C, at equilibrium,[20] which is easily accommodated within a combinatorial sample). The experimental space does, however, contain several $p_i$ axes, since S, Se, SnS and SnSe are all volatile. Together with the thermodynamic and kinetic variables $T$, $t$, $Q$ and $q$, this results in six axes that must be separately controlled in the experimental space for the pure compounds CZTS or CZTSe, and eight axes for the CZTSSe alloy.

Now consider an SDL configured according to **Figure 4**, producing repeatable combinatorial precursor libraries and feeding them into a dedicated reactor for defectome manipulation. The reactor provides independent control of S(e), Se(e), SnS(e), and SnSe(e) partial pressures, supports rapid thermal ramps, and is coupled to spatially-resolved Raman, photoluminescence, and four-point probe mapping. After establishing its baseline (Stage I), the SDL would constrain the experimental space by finding the single-phase boundary (Stage II). In **Table 4,** we estimate the resulting constrained ranges of the axes based on existing literature for the single-phase region, and suggest practical ranges for $t$, $Q$ and $q$, along with proposed *minimum* sampling resolutions.

In Stage III, the SDL would aim to fully explore the experimental space defined in **Table 4**. Even at minimal sampling resolutions, a gridded search of this space would constitute 11500 unique experiments for CZTS or CZTSe, and nearly 200000 for CZTSSe. Naturally, full sampling of these conditions is not necessary. In the SDL, active machine-learning algorithms can exploit smooth trends to achieve far more efficient sampling, using finer resolution where merited and sparser where changes are slow. Even so, for CZTSSe and other quaternary or pentanary materials systems, datasets comprising *hundreds to thousands of experiments* – each corresponding to synthesis and characterisation of one combinatorial sample, are likely required to achieve complete defectome mapping. In terms of operational time, if our SDL was capable of processing 20 combinatorial samples per day, it would take several months of continuous operation to generate such datasets. While this may seem extreme, it is important to realise that it is just proportionate to the complexity of the problem.

| Axis | Range of axis based on literature constraints | *Minimal* resolution | Total steps |
|---|---|---|---|
| $p_S$ and $p_{Se}$ | 6 orders of magnitude (temperature-dependent) [21] | 1 step per decade | 6 |
| $p_{SnS}$ and $p_{SnSe}$ | 3 orders of magnitude [21] | 1 step per decade | 3 |
| $T$ | 200-600°C. Upper limit depends on practical S pressure, lower limit on ordering kinetics[40] | 50°C steps | 8 |
| $t$ | 0.1-100 mins sufficient to probe kinetics of defectome changes over different length scales | 1 step per decade | 4 |
| $Q$ | 0.1-100°C/s plausible range for an RTP system | 1 step per decade | 4 |
| $q$ | 20-100% | 20% steps | 5 |

**Table 4:** Experimental axes, their ranges and minimum resolutions for defectome-aware exploration of CZTSSe in a scientific SDL.

**Coverage of existing datasets**

To underscore the potential value of SDL-based exploration, we contrast the dataset thus-far generated for CZTSSe – assuming that all published results were to be compiled – with the datasets that a scientific SDL of the type envisaged here would produce.

First, to assess how well experimentalists (this author included) have covered the experimental space described in **Table 4**, we performed a keyword-based survey of the CZTSSe literature. A Scopus query[§] targeting synthesis-focused studies since 2009 returned 1154 publications, of which 395 were directly downloadable. Full-text keyword analysis was carried out on this subset using a Python script based on the pypdf library (code available via GitHub[§§]), identifying reported experimental parameters using both keyword variants and unit-based searches.

Of the 395 papers analysed, 97% reported synthesis temperature and 91% reported synthesis time. By contrast, only 23% mentioned a sulfur or selenium pressure (across ~40 keyword variants), and just nine papers (2%) reported SnS or SnSe pressure. Cooling rate was mentioned in natural language in 10% of papers, but only six papers (1.5%) provided a numerical value.

Although this methodology is not perfect – for example, many studies may cite previously reported protocols – it still reveals a striking neglect of experimental axes that defect physics identifies as centrally important to synthesis-property relationships, especially partial pressures and cooling rates. This neglect is partly understandable: accurate control of S(e) and SnS(e) partial pressures is technically demanding, and equipment capable of fast thermal ramps and controlled cooling is not widespread. As a result, these parameters are often implicitly set by reactor design, not deliberately controlled and reported. Yet, this points to a clear need for improvement in synthesis equipment to make progress in future. In addition to the presence of these "hidden parameters", we can safely assume that the majority of publications are aiming at performance optimisation of CZTSSe solar cells, which means many experiments are likely to be clustered around known "good" regions of the parameter space. Meanwhile one can only imagine the many thousands of experiments going unreported due to "bad" results.

Overall, even if all data on CZTSSe synthesis from the past two decades could be compiled and harmonised, the issues noted above mean that the dataset would capture only a limited and ambiguous part of the overall synthesis-property relationships in the material. This is just one example of a domain-level problem that is repeated across functional materials research.

**The value of defect-aware SDL exploration**

We can contrast the above case with the datasets expected from defectome-aware exploration in an SDL. With reference to the Outputs in Fig. 4, the dataset produced will have several components: (1) the final surrogate models for Stages I-III of the exploration process; (2) metadata and full process data for each performed experiment, including measurement uncertainties; (3) extensive characterisation data for baseline material. To promote transferability and interpretability, it is encouraged to follow FAIR data principles and to include detailed information on sensors and measurement instruments in the SDL, such as type, calibration, geometry/location, etc. This can help determine the accuracy of reported values and judge the quality of the overall dataset.

The surrogate models for Stages II and III capture the synthesis-property relationships which the SDL is intended to map. Since they were derived by exploration, not only optimisation, and since their axes are directly related to physical phenomena, inspection of these models can reveal a wealth of information on the materials. From Stage II, the surrogate model structure corresponds to the fully delineated single-phase region across composition, partial pressure, temperature, and kinetic space. This constitutes an object for analysis unlike anything that presently exists. It defines the true domain in which intrinsic defect-mediated synthesis-property relationships operate, free from multiphase artefacts. It provides experimental access to chemical-potential space, enabling direct and quantitative connection between defect theory and real synthesis conditions. For applications, it defines processing windows and the true range of defect tunability in the pure compound. For theory, it imposes physical constraints on defect and kinetic models. The Stage II dataset, initially derived from functional property measurements, can be further validated by structure/phase measurements on a subset of samples, in particular for labelling the phases present just outside the single-phase region envelope.

From Stage III, the surrogate model (or models – one for each property measured) capture synthesis-property relationships across all dimensions of the single-phase region near and far from equilibrium, *and* in the multiphase vicinity captured on the combinatorial samples. Inspection of the surrogate models – particularly, extracting dominant gradients in properties - will reveal the action of all dominant defectome components, undergoing individual variations and having individual effects over various chemical potential regimes and timescales. While a given gradient in properties will not directly identify individual defectome components, it will impose very strong constraints on the possible candidates, by requiring consistency across the thermodynamic, kinetic, and multi-property trends. We could analyse the surrogate model mathematically to extract the partial derivative components of property changes, and use these to test any proposed models of the defectome constituents. We could also identify the regimes of synthesis space that avoid unwanted defect influences, which may suggest new synthesis pathways.

Compared to the patchy current data (using the example of CZTSSe), the described dataset would give theoreticians and experimentalists enormous support in planning and designing new investigations. Comparing datasets for different materials would also give valuable guidance in determining those which are most promising to investigate for applications, based on dataset-level metrics linked to synthesisability (i.e., easy kinetics) and defect tolerance (low sensitivity to perturbations).

Such datasets could be generated automatically in exactly the same way for any multinary inorganic film – whether for optoelectronics or other functional application, without prior art. Via accumulating multiple examples, initially for related materials and later for increasingly diverse ones, the SDL procedures will be refined to accelerate the data-collection campaign. Over time, we predict that the accumulation of such rich datasets in public/community databases such as NOMAD[41] or HTEM[42], will be the key to training powerful, synthesis-aware AI models for designing new materials in diverse contexts.

**Our SDL**

Overall, we feel that there is a strong case for realising the type of scientific SDL envisaged here, although this is not without challenges. At Uppsala university, a platform is taking shape that we hope will demonstrate at least some of the capabilities envisaged. The system, nicknamed BERTHA, is based on a sputtering chamber, coupled via a glovebox to a rapid thermal processor (RTP). Production of combinatorial samples and defectome evolution experiments take place in a hybrid two-stage approach as follows: First, the non-volatile (metal) components of the material in question are sputtered as a combinatorial thin film. Up to six elements can be combined, with the composition distribution being inferred using in-situ measurements coupled to machine-learned models of each source.[43] Next, the film is transferred to the RTP, a high-vacuum system which is provided with an $H_2S$ supply. Here, the sample is first reacted with $H_2S$ to complete the baseline formation (synthesis step 1 in the scheme of **Figure 4**), whereupon the conditions are immediately adjusted to initiate a specific defectome control experiment (synthesis step 2). The main advantage of this route is that it allows considerably faster turnover than other PVD options: metal precursors can be sputtered at high rates, requiring only minutes to generate sufficient film thickness. The pumping time required during load/unload cycles means vacuum systems are often considered too slow for SDLs. This is addressed both by the elimination of humidity due to the glovebox, and by a load-lock on the sputter chamber. As a result, the overall sample turnover can be as fast as 20 mins (depending on RTP duration), enabling in principle 20-30 combinatorial samples to be produced per day. Automation of sample transfer is to be achieved by a robot arm placed in the glovebox. More details of this system, automation of baseline process development using optical-only measurements, and the design of SDL algorithms for characterising single-phase regions, will be presented in future publications.

# Conclusions

We have argued that correctly-designed and operated self-driving laboratories (SDLs) can serve not only as optimisation engines, but as scientific instruments, generating physically grounded datasets capable of advancing mechanistic understanding, and thus becoming a cornerstone of data-driven science. In complex multinary inorganic materials such as novel optoelectronics, synthesis-property relationships are governed primarily by defect physics and chemistry. Because defect populations and their spatial organisation cannot generally be resolved directly – particularly in high-throughput settings – progress depends on constructing datasets in which thermodynamic and kinetic perturbations are systematically applied and defect-sensitive properties are measured in parallel. We propose that this should be the main task for exploratory SDLs in the field of optoelectronics and related functional materials.

To frame this challenge, we introduced the concept of the *defectome* as a collective descriptor of defect populations and configurations across length scales. While not intended as a definitive mathematical formalism, the defectome provides a useful abstraction for reasoning about the intermediate state linking synthesis conditions to functional properties. From this perspective, we articulated four design principles for exploration-driven SDLs: (i) combinatorial thin films as the unit experiment, (ii) separation of phase formation from defectome manipulation, (iii) explicit parameterisation of the thermochemical and kinetic experimental space, and (iv) use of the single-phase region to constrain exploration. Applied to the example of $Cu_2ZnSn(S,Se)_4$, this framework reveals that the experimentally relevant parameter space spans multiple largely unexplored axes, explaining in part the limited mechanistic clarity achieved despite decades of study, and showing the potential for a better approach.

The central claim of this work is therefore methodological rather than materials-specific: that deliberately designed, synthesis-aware SDLs can generate datasets with sufficient breadth and structure to constrain defect models in ways not accessible to either first-principles calculations or conventional experimentation alone. Realising such systems will require careful integration of hardware design, sensing, process modelling, and physics-informed ML. However, the potential payoff is substantial: a pathway toward transferable, synthesis-aware materials models and datasets and, ultimately, rational materials design grounded in experimentally resolved defect physics.

# Data availability

Data and code for literature analysis are freely available via GitHub https://github.com/Automate-Solar/Faraday

# Conflicts of interest

There are no conflicts to declare

# Acknowledgements

The author gratefully acknowledges financial support by the Wallenberg Initiative Materials Science for Sustainability (WISE) funded by the Knut and Alice Wallenberg Foundation, from the Swedish Research Council and from the AI4Research initiative at Uppsala University.

## Notes and references